\documentclass[journal=jacsat,manuscript=article]{achemso}

\usepackage{amsmath,amssymb,amsfonts}
\usepackage{bm}
\usepackage{color, soul}
\usepackage{graphicx}
\usepackage{float}
\usepackage{wrapfig}
\usepackage{verbatim}
\usepackage[english]{babel}
\usepackage[dvipsnames]{xcolor}
\usepackage[
    colorlinks=true,
    bookmarks=false,
    citecolor=blue,
    urlcolor=blue,
    linkcolor=blue,
]{hyperref}
\usepackage{natbib}
\usepackage{subfigure}
\usepackage{amstext, mathrsfs, textcomp}
\usepackage{multirow}
\usepackage{siunitx}
\usepackage{enumerate}
\usepackage{enumitem}
\usepackage{epstopdf}
\usepackage{tabularx}
\makeatletter
\usepackage{lineno}

\usepackage[version=3]{mhchem} 
\usepackage{xr}
\usepackage{colortbl}

\author{N. Glebov}
\affiliation[ITMO]
{School of Physics and Engineering, ITMO University, St. Petersburg 197101, Russia}
\alsoaffiliation[EPFL]
{Institute of Bioengineering, École Polytechnique Fédérale de Lausanne, Lausanne 1015, Switzerland}

\author{M. Masharin}
\affiliation[EPFL]
{Institute of Bioengineering, École Polytechnique Fédérale de Lausanne, Lausanne 1015, Switzerland}
\author{A. Yulin}
\affiliation[ITMO]
{School of Physics and Engineering, ITMO University, St. Petersburg 197101, Russia}
\author{A. Mikhin}
\affiliation[ITMO]
{School of Physics and Engineering, ITMO University, St. Petersburg 197101, Russia}
\author{M. R. Miah}
\affiliation[UNAM]
{Department of Electrical and Electronics Engineering, Department of Physics, UNAM - Institute of Materials Science and Nanotechnology and The National Nanotechnology Research Center, Bilkent University, Ankara 06800, Turkey}
\author{H. V. Demir}
\affiliation[UNAM]
{Department of Electrical and Electronics Engineering, Department of Physics, UNAM - Institute of Materials Science and Nanotechnology and The National Nanotechnology Research Center, Bilkent University, Ankara 06800, Turkey}
\alsoaffiliation[NTU]{ Luminous! Center of Excellence for Semiconductor Lighting and Displays School of Electrical and Electronic Engineering, School of Physical and Mathematical Sciences, and School of Materials Science and Engineering, Nanyang Technological University, Singapore 639798, Singapore}
\author{D. Krizhanovskii}
\affiliation[US]
{Department of Physics and Astronomy, University of Sheffield, Sheffield S3 7RH, United Kingdom}
\author{V. Kravtsov}
\affiliation[ITMO]
{School of Physics and Engineering, ITMO University, St. Petersburg 197101, Russia}
\author{A. Samusev}
\affiliation[TUD]
{Experimentelle Physik 2, Technische Universität Dortmund, 44227 Dortmund, Germany}
\email{anton.samusev@gmail.com}
\author{S. Makarov}
\affiliation[ITMO]
{School of Physics and Engineering, ITMO University, St. Petersburg 197101, Russia}
\alsoaffiliation[HEU]
{Qingdao Innovation and Development Center, Harbin Engineering University, Qingdao 266000 Shandong, China }

\email{s.makarov@metalab.ifmo.ru}

\title[An \textsf{achemso} demo]
{Room-temperature exciton-polariton-driven self-phase modulation in planar perovskite waveguides}

\abbreviations{IR,NMR,UV}

\begin{document}

\begin{abstract}
Optical nonlinearities are crucial for advanced photonic technologies since they allow photons to be managed by photons. Exciton-polaritons resulting from strong light-matter coupling are hybrid in nature: they combine small mass and high coherence of photons with strong nonlinearity enabled by excitons, making them ideal for ultrafast all-optical manipulations. Among the most prospective polaritonic materials are halide perovskites since they require neither cryogenic temperatures nor expensive fabrication techniques. Here we study strikingly nonlinear self-action of ultrashort polaritonic pulses propagating in planar MAPbBr$_3$ perovskite slab waveguides. Tuning input pulse energy and central frequency, we experimentally observe various scenarios of its nonlinear evolution in the spectral domain, which include peak shifts, narrowing, or splitting driven by self-phase modulation, group velocity dispersion, and self-steepening. The theoretical model provides complementary temporal traces of pulse propagation and reveals the transition from the birth of a doublet of optical solitons to the formation of a shock wave, both supported by the system. Our results represent an important step in ultrafast nonlinear on-chip polaritonics in perovskite-based systems. 
\end{abstract}


\section{Introduction}
One of the most important tasks of modern nanophotonics is leveraging nonlinear optical effects for efficient light manipulation at the nanoscale, enabling the development of ultrafast, compact photonic devices and circuitry. Particular applications of nonlinear nanophotonics include ultrafast signal processing and full optical switches\cite{chai2017ultrafast, ballarini2013all}, supercontinuum light generation\cite{dudley2006supercontinuum, walker2019spatiotemporal} or ultrashort pulse generation\cite{zhao2012ultra}. Nonlinear effects arise from photon-photon interactions that are mediated by materials. These effects are naturally weak because they depend on the nonlinear response of the material to electromagnetic fields\cite{kauranen2012nonlinear}. However, they can be significantly strengthened in environments where the electromagnetic field and photon-photon interaction are enhanced by specific material mechanisms, such as the formation of exciton-polaritons in semiconducting structures\cite{RevModPhys.71.1591}.

Exciton-polaritons are bosonic quasiparticles formed through the strong coupling of light and matter, garnering significant attention due to their unique properties\cite{sanvitto2012exciton}. The photonic component of exciton-polaritons offers high group velocity\cite{walker2013exciton}, low effective mass and extended coherence times\cite{deng2010exciton, sanvitto2016road}, while the excitonic component introduces strong nonlinearity driven by Coulomb interactions between the quasiparticles, which can significantly enhance the corresponding nonlinear optical response\cite{PhysRevB.59.10830, masharin2022polaron, suarez2021enhancement}.  The Kerr nonlinearity in such systems can be 3-4 orders of magnitude stronger than in typical semiconductor or dielectric materials, where light couples only weakly to crystal excitations\cite{walker2013exciton}.

One of the most extensively studied and commonly utilized material platforms for exciton-polariton systems is the GaAs quantum well (QW)\cite{weisbuch1992observation,deng2002condensation,sanvitto2016road}. Despite the huge progress in this field recently, this platform has low exciton binding energy, which limits it to cryogenic temperatures\cite{walker2015ultra, kasprzak2006bose}. This temperature constraint can be addressed using wide-bandgap semiconductor QWs like ZnO\cite{van2006exciton} or GaN\cite{semond2005strong}, although usually, these materials require complex and costly fabrication techniques, such as epitaxial growth. Recently, monolayer transition metal dichalcogenides have emerged as promising candidates for room-temperature polariton systems\cite{stepanov2021exciton}, yet their practical applications remain limited by challenges in scalable manufacturing.

Planar systems such as metasurfaces and waveguides are among the most promising platforms for on-chip communication and data processing nowadays. However, the nonlinear response of polaritonic guided systems at room temperature remains largely unexplored, even though these systems have the potential to exhibit various nonlinear phenomena, including solitons\cite{walker2015ultra} and pulse modulation\cite{benimetskiy2023nonlinear, di2021ultrafast}.  Halide perovskites exhibit remarkable excitonic properties, making room-temperature applications much more feasible, and offer the advantage of easy and cost-efficient fabrication\cite{masharin2023room, Tonkaev2023May, makarov2019halide}. Perovskites have recently shown their potential for advanced applications, as demonstrated by successful room-temperature optical switching\cite{su2021optical} and edge lasing from a crystalline waveguide\cite{kkedziora2024predesigned}. However, the vertical geometry remains a drawback when designing photonic circuits using perovskites.

In this work, we study the nonlinear propagation of the guided exciton-polariton pulses in a planar perovskite-based waveguiding system at room temperature. We inject (extract) femtosecond pulses into (from) the waveguide using resonant grating couplers imprinted in the perovskite waveguide. As the fluence of the input pulse increases, the spectral shape of the pulse transforms during its propagation through the waveguide due to the interplay of self-phase modulation (SPM), group velocity dispersion (GVD), waveguide dispersion, influence of the exciton reservoir, and other processes. Self-phase modulation is a nonlinear optical effect where a pulse’s phase shifts due to intensity-dependent changes in the medium’s refractive index, broadening the pulse spectrum. Microscopically, this effect arises from polariton-polariton scattering, where polariton pairs interact and redistribute energy across different frequencies, effectively modulating the pulse phase as it propagates. The experimental results are supported by numerical simulations of the studied system. Our numerical analysis further reveals that SPM induces the temporal splitting of the input pulse into two separate pulses, influenced by strong spectrally inhomogeneous dispersion and nonlinearity inherent to the polaritonic nature of the system. Our experimental findings, showing wavelength- and distance-dependent femtosecond pulse evolution, align well with the theoretical model, allowing us to explicitly interpret the observed phenomena.

\label{ch:results}
\section{Experiment}

\begin{figure}[h!]
    \centering
    \includegraphics[width=0.8\linewidth]{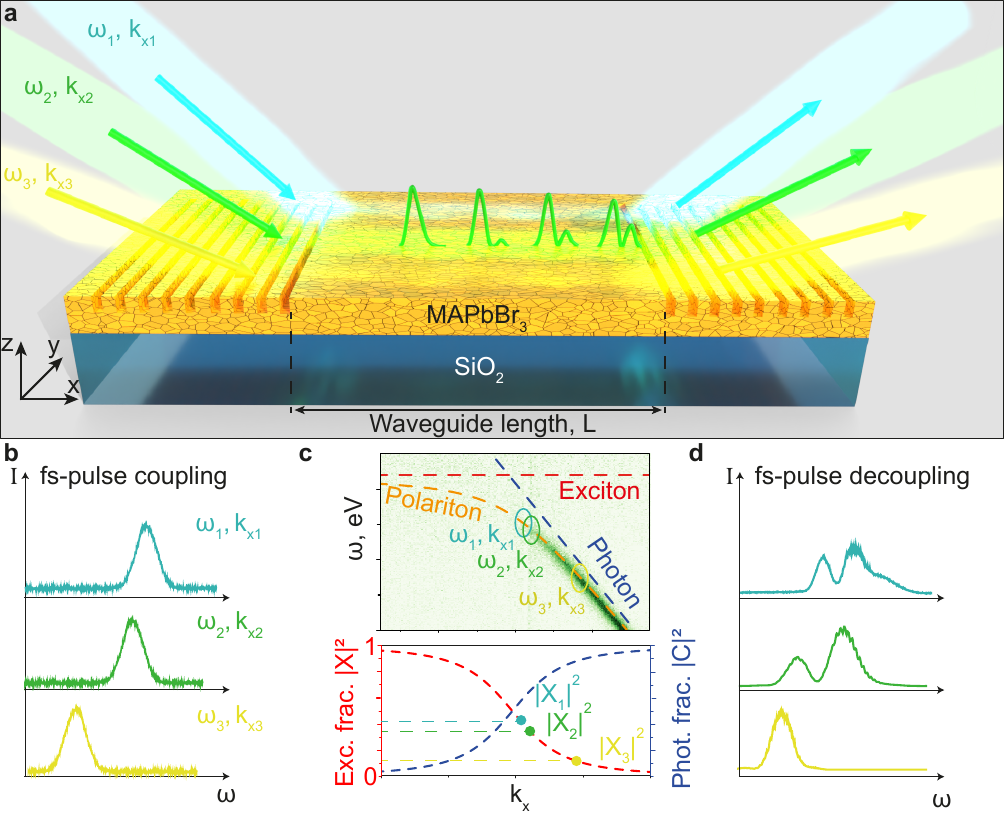}
    \caption{\textbf{Propagation of exciton-polariton wavepackets in a halide perovskite waveguide.} (\textbf{a}) Sketch of the studied system, demonstrating the processes of coupling/decoupling of light into/from the waveguide for three different detunings. The spectrum of the propagating pulse undergoes self-phase modulation and nonlinear group velocity dispersion, as illustrated schematically on the waveguide surface. (\textbf{b}) Spectra of femtosecond pulses incident on the coupler for three different detunings.(\textbf{c}) Top: Measured angle-resolved transmittance spectrum from decoupler with excitation by white light. The red and dark blue dashed lines correspond to uncoupled exciton and photon. The Orange dashed line is polariton.  Ellipses show polariton states with particular energies and wavevectors excited resonantly using a femtosecond laser. Bottom: Calculated Hopfield coefficients as a function of the wavenumber. (\textbf{d}) Spectra of femtosecond pulses transmitted through the waveguide for three different detunings. }
    \label{fig:1}
\end{figure}

In our experiments, we fabricate planar waveguides based on MAPbBr$_3$ thin films (see schematic illustration in \autoref{fig:1}a). One of the key advantages of halide perovskites is that they can be synthesized through solution-based methods, enabling low-cost fabrication and making them highly promising for practical applications\cite{jeon2014solvent}. Due to the relatively high refractive index of 2.1-2.3 in the 500-600~nm wavelength range\cite{refractiveindexMaPbBr3} a perovskite thin film works as a planar waveguide with sub-wavelength thickness. Due to its sub-light-line dispersion, a waveguide mode cannot be excited directly in free space and requires a coupling mechanism to match the necessary wavevector\cite{snyder1983optical}. One of the most effective and straightforward methods to excite waveguide modes is through the use of periodic structures such as grating couplers.\cite{cheben2018subwavelength}. Various techniques can create periodic structures on halide perovskite surfaces, including nanoimprint lithography\cite{makarov2017multifold,masharin2023AFM}, mechanical scanning probe lithography\cite{glebov2023mechanical}, and direct laser writing\cite{zhizhchenko2021direct}. We choose nanoimprint lithography for its high resolution, direct pattern transfer, and compatibility with the soft crystal lattice of halide perovskites, making it particularly well-suited for structuring these materials.

We employ the solution-based method via spin coating and nanoimprinting to fabricate planar waveguides with elements for coupling to the waveguide mode (see Methods). Glass molds are prepared by electron beam lithography, following established protocols (see Methods). The final system comprises three key elements: a grating coupler, an unstructured planar waveguide, and an identical grating decoupler. Several samples with different waveguide lengths have been fabricated and studied with distances between couplers and decouplers ranging from 50 to 200~$\mu$m. The thickness of the waveguide is 110~nm defined by the thickness of the MAPbBr$_3$ thin film. The gratings have a period of 280~nm, a modulation depth of 65~nm, and a fill factor of 0.48. The topography of the fabricated waveguide sample, as measured by atomic force microscopy (AFM), is shown in Supplementary Information (SI) (Figure S2a,c). All geometrical parameters are selected based on numerical simulations to achieve waveguide modes near the exciton resonance of our material ($\sim 2.32$~eV).

We perform measurements using a custom back-focal plane setup for angle-resolved transmission and reflection spectroscopy paired with a wavelength tunable femtosecond laser and a halogen lamp for the excitation as schematically illustrated in Figure S3a,b (see Methods for details). In order to estimate the exciton-photon coupling strength in the perovskite waveguide, we first obtain the angle-resolved reflectance spectra by measuring the leaky modes of couplers. The experimental results (Figure S4a) show the dispersion of the waveguide leaky modes in the reflection.  Dispersion of the waveguide leaky modes behavior demonstrates a strong curvature near the exciton resonance, which is typical anticrossing behavior in strong-coupled systems\cite{deng2010exciton}.

We fit the measured leaky modes with the two-coupled oscillators model\cite{deng2010exciton, hopfield1958theory}(see SI for details).
The Rabi splitting and coupling constant are estimated with this model as $\Omega_R$=104~meV and $g = 54$~meV, respectively.
The estimated Rabi splitting and coupling constant satisfy the criteria of the light-matter strong coupling at room temperature as $\Omega_R>(\gamma_{ex} + \gamma_{ph})/2$ and $g>|\gamma_{ex} - \gamma_{ph}|/2$,\cite{schneider2018two,kondratyev2023probing} since the estimated losses of uncoupled photons and excitons are $\gamma_{ph} = 16$~meV and $\gamma_{ex} = 14$~meV, respectively. 

Transmission through the waveguide is measured by coupling incident TE-polarized white light (with the electric field parallel to the coupler grooves) into the input coupler and collecting the all output signal from the decoupler (as shown in the inset of Figure S3b for waveguide length of 50~$\mu$m). As a result, we obtain the angle-resolved linear transmittance spectrum of the waveguide mode (Figure S4b). We can observe only one branch of the dispersion with group velocity directed towards the decoupler, where the signal is collected. The spectral region of the dispersion close to the exciton level has low transmission due to exciton losses. 

The nonlinear response of the exciton-polariton waveguide system is studied utilizing the wavelength-tunable femtosecond laser pulses. Transmission of the laser pulses is measured using the same method mentioned above for the white light (\autoref{fig:1}a).  The incident angle and wavelength of the excitation laser are adjusted to achieve resonant pumping (\autoref{fig:1}b,c), as detailed in the Methods section. We study the waveguide transmission as a function of the input fluence from resonant femtosecond pumping at various waveguide lengths ($L$) and detunings ($\delta$) from the exciton resonance energy. Studied polariton states are demonstrated on the waveguide polariton dispersion by ellipses on the transmittance angle-resolved spectrum (\autoref{fig:1}c). We consider three different detunings from the exciton resonance of $-64, -74, -146$~meV, which correspond to Hopfield coefficients of excitonic fractions\cite{hopfield1958theory, masharin2023room} of $|X_1|^2 = 0.47$, $|X_2|^2 = 0.41$, $|X_3|^2 = 0.15$ calculated as outlined in SI. Laser pulse transmittance spectra, corresponding to the mentioned detunings are shown in \autoref{fig:1}d.

\begin{figure}[h!]
    \includegraphics[width=1\linewidth]{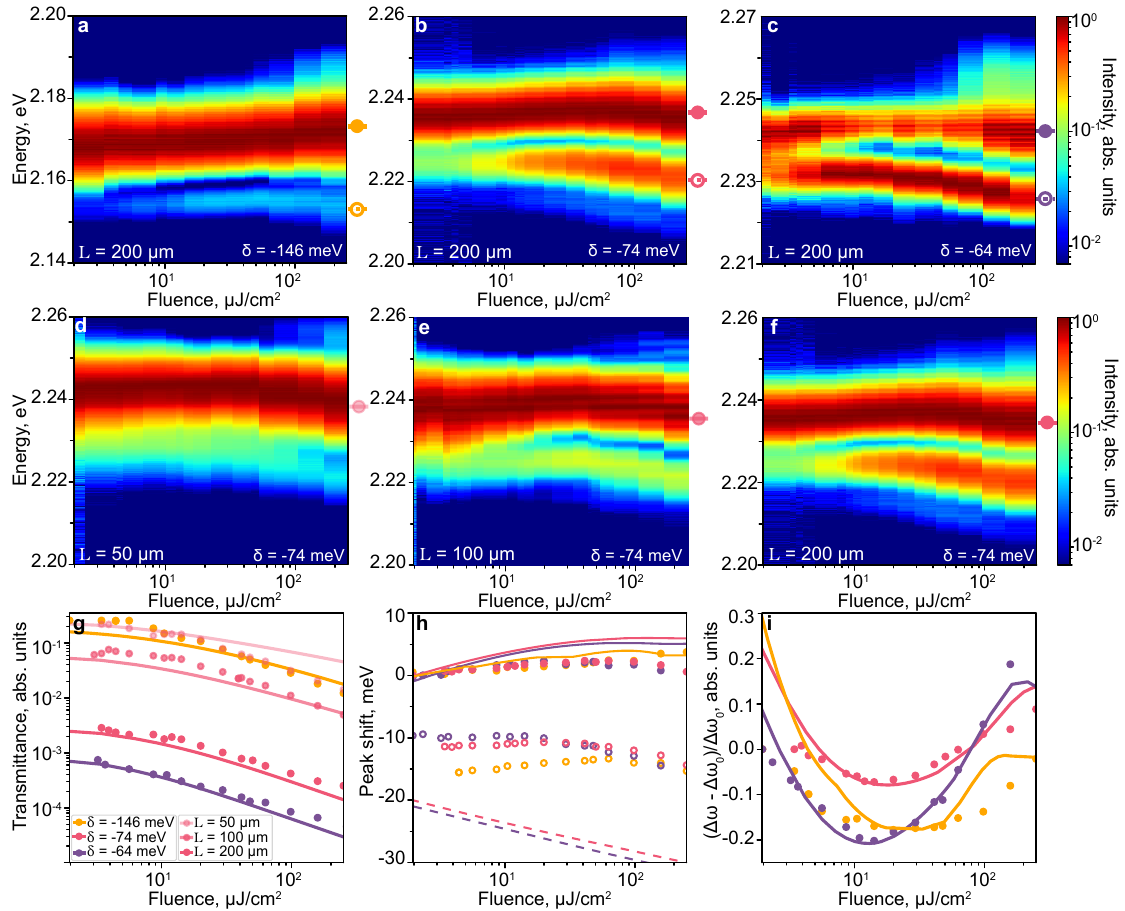}
    \caption{ \textbf{Experimental results on laser pulse propagation in a perovskite planar waveguide in different spectral ranges.}  Normalized transmitted pulse spectra vs. input fluence for (\textbf{a-c}) three different detunings from the exciton level and waveguide length $L = 200 \ \mu$m and (\textbf{d-e}) three different waveguide lengths and detuning  $\delta = -74$~meV. (\textbf{g}) The transmittance of the waveguide normalized on the pump fluence adjusted for the effectiveness of the coupler for different detunings and waveguide lengths.  (\textbf{h}) Peak shift from the initial position at small fluence for two peaks (high energy peak - filled circles, low energy peak - empty circles). (\textbf{i}) The relative spectral width of the pulse normalized on the initial spectral width at the lowest fluence for different detunings. The lines correspond to the theoretical calculations. }
    \label{fig:2}
\end{figure}

In the experiment, we measure the transmitted spectra of the fs-laser pulse across various detunings and waveguide lengths, presenting the normalized intensity spectra as a function of input fluence in \autoref{fig:2}a-c for different detunings and \autoref{fig:2}d-f for different waveguide lengths. We generally observe spectral broadening of the output pulses with increasing pump fluence and the emergence of an additional peak. This effect is consistently observed across all energy detunings ({\autoref{fig:2}}a-c) and waveguide lengths ({\autoref{fig:2}}d-f). We observe the occurrence of additional peak with lower energy than the energy of the input pulse. 

At a fixed waveguide length of 200~$\mu$m, the increasing of the excitation detuning leads to the attenuation of the second peak formation in the output pulse spectra ({\autoref{fig:2}}a-c).  At the maximal detuning of -146~meV, the low-energy peak emerges at the input fluence of approximately 10~$\mu$J/cm$^2$, with a notably smaller amplitude compared to the main peak (\autoref{fig:2}a). As the pump fluence increases, the positions of the peaks undergo a blueshift across the entire examined range. Additionally, the spectral width behaves non-monotonically with fluence: it initially decreases until about 40~$\mu$J/cm$^2$, and then starts to increase. At the minimal detuning of -64~meV, the low-energy peak occurs with significant amplitude at the input fluence of approximately 4~$\mu$J/cm$^2$ (\autoref{fig:2}c). With increasing pump fluence, the positions of both peaks initially experience a blueshift, followed by a redshift. The spectral width shows a similar trend to that observed at larger detunings, with the narrowing to broadening crossover at approximately 10~$\mu$J/cm$^2$. 

At a fixed detuning of -74~meV, the increasing propagation length causes the formation of a second peak in the spectra of the output pulse ({\autoref{fig:2}}d-f). The generation of new frequency components occurs continuously along the propagation in the waveguide. The amplitude of the low-energy peak for a waveguide length of 50~$\mu$m (\autoref{fig:2}d) is so small within this fluence range that it is nearly indistinguishable from the main peak. However, in the longer waveguide of 200~$\mu$m, the low-energy peak becomes significantly more pronounced (\autoref{fig:2}f). Additionally, as the waveguide length increases, the splitting between the low-energy and high-energy peaks begins to occur at progressively lower fluences. 



The transmittance coefficient of the waveguide, determined by energy integration of the spectra and normalized on fluence for different waveguide lengths and different detunings, is plotted in \autoref{fig:2}g (filled circles). We estimate coupling efficiency and account for input and output pulse energy in the experiment (see SI for details). Across all detunings and waveguide lengths, transmittance decreases nonlinearly as fluence increases; in this nonlinear regime, both transmittance and pulse shape are affected by the input fluence (see SI for linear regime details). 

In order to describe the evolution of the pulses in the waveguide with the increase of incident fluence, we estimate the central frequency, linewidth, and amplitude of each peak (see SI for details). \autoref{fig:2}h (circles) shows the pump flunce-dependent shift of the peak position. In the case of the largest detuning of -146~meV used in the experiment, the high-energy peak exhibits only a blueshift, while the low-energy peak shows both a blueshift and a subsequent redshift. At the other two detunings, -74~meV and -64~meV, the high-energy peak exhibits a blueshift followed by a redshift, while the low-energy peak shows a redshift. {\autoref{fig:2}}i (circles) demonstrates the relative change in spectral linewidth with fluence, normilized to the low-fluence value. The spectral linewidth initially narrows and then broadens  with fluence across all detunings and waveguide lengths.

\section{Theory}
To gain an insight into the mechanisms governing the experimentally observed evolution of the propagating through the waveguide laser pulses with increasing incident fluence, we develop a theoretical model. It incorporates the strong coupling between excitons and photons,  exciton-exciton repulsion\cite{masharin2023room}, and interaction of quasiparticles with an incoherent exciton reservoir, yielding the following system of equations:\cite{fieramosca2019two,wu2021nonlinear,masharin2024giant,wouters2007excitations, skryabin2017backward}
\begin{equation}
    \begin{cases}
        \partial_tA + v_g \partial_xA = -\gamma_{ph}A + i\Omega_R\psi + f_p(x,t),\\
        \partial_t\psi = -(\gamma_{ex} + \gamma_{res})\psi + i\Delta\psi -\Gamma\psi + i\Omega_R A, \label{coh_exciton_eq}\\
        \partial_t\rho  = -\gamma_{\rho}\rho + 2\gamma_{res}|\psi|^2
    \end{cases}
\end{equation}

The first equation is written for the slowly varying amplitude $A$ of the photon-guided mode.  We approximate photonic dispersion in the studied range of wavevectors by a linear function, with group velocity $v_g$ evaluated at the exciton resonance frequency and dissipation rate $\gamma_{ph}$. The photonic component is excited by the driving force (currents) $f_p$ produced by the incident pulse in the area of the input coupler (see SI for details).

The photonic subsystem is coupled to the excitonic one with the coupling strength characterized by Rabi splitting $\Omega_R$.  The coherent excitons are described by the slowly varying envelope amplitude $\psi$. The losses in the exciton sub-system are defined by the total dissipation rate $\gamma_{ex}+\gamma_{res}$ consisting of the coherent exciton decay rate $\gamma_{ex}$ and the rate of the scattering to the reservoir of the incoherent excitons $\gamma_{res}$ (incoherent exciton density). The shift of the coherent exciton frequency due to nonlinear effects is accounted by $\Delta$ which is a function of $|\psi|^2$ and $\rho$. Describing the dynamics in terms of slowly varying amplitudes, we choose the carrier frequency to be equal to the exciton resonance frequency in the linear regime, so $\Delta(|\psi|^2=0, \rho=0)=0$.

In the strong exciton-photon coupling regime, excitonic and photonic dispersions anticross, and the resulting exciton-polaritons acquire strong dispersion near the resonance. The dispersion of these hybrid excitations (polaritons) is given by
\begin{equation}
\omega(k)=\frac{ v_g k+i(\gamma_{ph}+\gamma_{ex}+\gamma_{res}) \pm\sqrt{\left(v_g k+i(\gamma_{ph}-\gamma_{ex}-\gamma_{res}) \right)^2+4\Omega_R^2  }   }{2}. \label{dispersion}
\end{equation}

To reproduce the dynamics observed in the experiments, it is necessary to account for the reservoir of long-living incoherent excitons characterized by density $\rho$ and lifetime $\gamma_{\rho}$. We assume uni-directional scattering from coherent to incoherent excitons with rate $\gamma_{res}$. The reservoir may form through scattering into excitonic states within the tail of the inhomogeneously broadened exciton line\cite{whittaker1998determines, vinattieri1994exciton}, where the density of states is significantly higher than for polaritons. Alternatively, the reservoir can consist of localized excitons, indirect excitons, or other dark excitonic states.\cite{KRIZHANOVSKII2001435, sarkar2010polarization,menard2014revealing,tollerud2016revealing}

While we assume the photonic component to behave linearly with fluence, for excitons both the resonance frequency and decay rate are density-dependent. These nonlinear dependencies of the resonant frequency and the losses are accounted for by exciton density-dependent parameters $\Delta$ and $\Gamma$ (see SI for details).


\begin{figure}[H]
\centering
\includegraphics[width=1\linewidth]{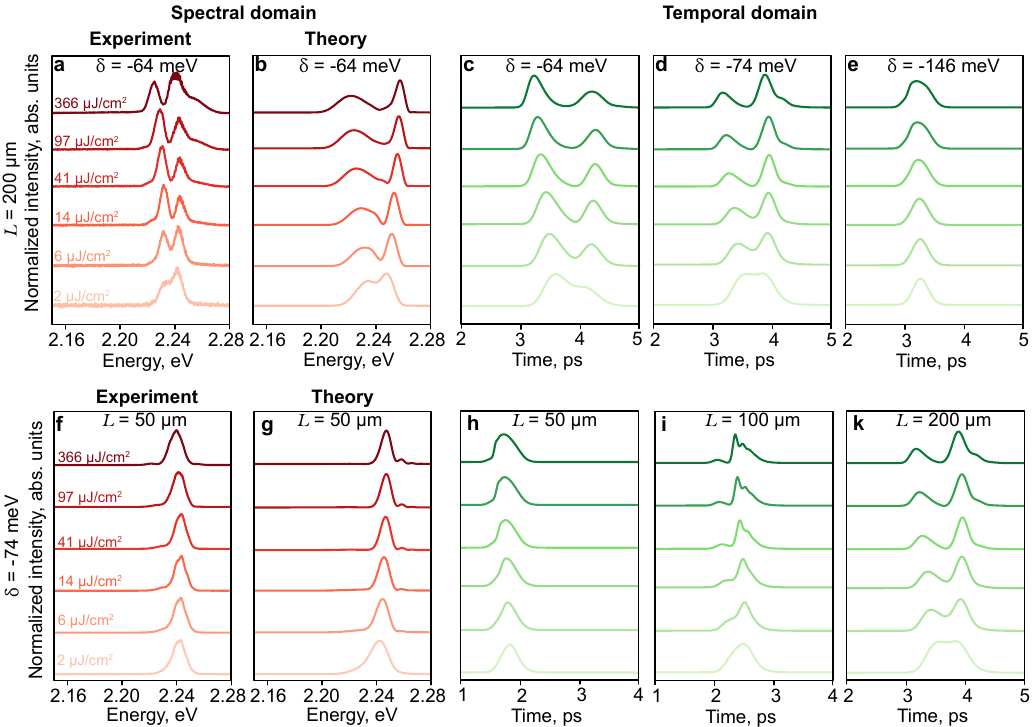}
\caption{\textbf{Theoretical modeling of laser pulse propagation in a perovskite planar waveguide.} Normalized transmitted spectra plotted in the dependence of input fluence for the detuning of -64 meV and waveguide length 200 $\mu$m and for the detuning of -74 meV and waveguide length 50 $\mu$m (\textbf{a, f}) for the experiment and (\textbf{b,g}) theoretical calculation, respectively.  Simulated transmitted temporal pulse profiles normalized on input pulse plotted as the dependence on input fluence for (\textbf{c-e}) different detunings and (\textbf{h-k}) waveguide lengths.}
\label{fig:3}
\end{figure}

\section{Discussion}
We numerically solve the system of equations described above using the split-step method, calculating the photonic field as a function of both time and position in the waveguide.  The model gives good qualitative and quantitative agreements with the experimental data (\autoref{fig:2}g-i (lines) and \autoref{fig:3}a-b,f-g). 
The transmittance behavior observed in the experiment and the model exhibits the same pattern: an initial linear regime (the transmittance does not depend on input fluence, see SI) followed by a nonlinear decrease with rising input fluence  (\autoref{fig:2}g). This phenomenon occurs because nonlinear losses outweigh the effect of exciton resonance blueshift. In the case of fluence-dependent exciton resonance blueshift, the polariton losses at the same energy decrease -- since these losses are primarily determined by the excitonic component. As a result, transmittance increases\cite{benimetskiy2023nonlinear}.  However, in $\text{MAPbBr}3$, nonlinear losses quickly dominate over the blueshift effect, leading to a net decrease in transmittance. Additionally, the overall transmittance decreases for longer waveguides or detunings corresponding to higher exciton fractions (\autoref{fig:2}g). With reduced detuning from the exciton resonance, transmittance declines further due to increased losses, as the exciton fraction—and thus polariton loss—rises. Near the exciton resonance, effective losses $\gamma_{eff}$ increase, while the group velocity $v_{g}$ decreases, jointly reducing waveguide transmittance.

Increasing the input pulse energy blueshifts the exciton resonance, reducing the difference in absorption rates across frequencies. This decrease in absorption variation lowers the redshift of the output pulse at higher pump powers, leading the central frequency of the output spectrum to shift towards the blue side of the spectrum. Changes in peak position are governed by two parameters related to shifts in the central exciton frequency, $\alpha_3$ and $\alpha_5$, corresponding to 2-particle and 3-particle interactions as discussed in SI. The parameter $\alpha_3$ is positive, contributing to an increase in the exciton frequency (a blueshift), while $\alpha_5$ is negative, leading to a saturation in the exciton frequency shift\cite{masharin2023room}. As the input fluence increases, $\alpha_3$ initially dominates, followed by a significant contribution from $\alpha_5$. Consequently, both the experimental and numerical data reveal an initial blueshift followed by the shift saturation as the input fluence increases (\autoref{fig:2}h). Once the blueshift reaches saturation, a redshift occurs as incoherent excitons absorb higher frequencies, effectively shifting the pulse toward the red side of the spectrum. 

In both theoretical and experimental data, the spectral separation between the two peaks widens with increasing fluence. The high-energy peaks are well-captured by the model, displaying an initial blueshift followed by a redshift, consistent with experimental observations (\autoref{fig:2}h). For the low-energy peaks, the model predicts only redshifts, without the initial blueshift observed experimentally. This discrepancy may arise because the model accounts only for one-dimensional propagation and does not fully accommodate variations in coupling strength at different detunings, along with minor mismatches in input energy. Additionally, slight variations in the input pulse shape, which cannot be directly measured, are expected.

 {\autoref{fig:2}}i demonstrates the relative change in spectral linewidth compared to the initial linewidth at low fluence, as the fluence varies. The spectral peak initially narrows and then broadens across all detunings and waveguide lengths. This behavior of the pulse width is common for the solitonic regime of propagation\cite{walker2015ultra}. Weak linear losses lead to spectral compression by adding a perturbative term to the Nonlinear Schrödinger Equation (NSE). As the pulse loses energy, its peak power and nonlinear phase shift decrease, necessitating a narrower spectrum to balance nonlinear and dispersion lengths. There is strong agreement between the experiment and the simulation, with discrepancies only appearing at high fluences, where higher-order nonlinearities may need to be incorporated into the model. 
 
  \autoref{fig:3} shows normalized transmitted spectra plotted in the dependence of input fluence for different detunings and waveguide lengths, alongside a comparison with numerical data. The experimental spectra presented in \autoref{fig:3}a for $\delta = -64$~meV and $L = 200 \ \mu$m  and \autoref{fig:3}f  for $\delta = -74$~meV and $L = 50 \ \mu$m are in good agreement with our calculated result in \autoref{fig:3}b,g. Both the experimental and theoretical results show two distinct spectral peaks with similar trends in central frequency shifts and linewidth. In the experiment, the pulse structure appears more symmetric, resembling classical self-phase modulation, while in the model, the shape of the low-energy peak shows slight variation. This difference may result from minor variations in the input pulse shape.

We propose the presence of two key nonlinear effects in the perovskite waveguide under strong light-matter coupling: interplay between self-phase modulation (SPM) and nonlinear group velocity dispersion (GVD), and self-steepening, both of which play crucial roles in shaping the output pulse dynamics. In contrast to traditional fiber systems\cite{dudley2006supercontinuum}, the nonlinear processes here are associated with exciton-polariton dispersion and its nonlinear behavior. The first process, balancing of SPM and GVD, leads to soliton formation \cite{agrawal2000nonlinear, kivshar2003optical}. This effect symmetrizes the pulse in the time domain, countering the inherent asymmetry caused by dispersion. In a linear regime, output pulses tend to be distorted due to dispersion effects, but the soliton regime restores symmetry by balancing dispersion and nonlinearity \cite{walker2015ultra,yulin2022bright}. The second process is the so-called self-steepening effect whose physical origin is the dependence of the group velocity on the local intensity of the field. The self-steepening results in the formation of shock waves \cite{demartini1967self}. This phenomenon is well known in fiber optics \cite{Agrawal2012NonlinearFO} and consists in the formation of a sharp leading edge followed by a slowly decaying tail. 

One of the main advantages of exciton-polariton systems is the potential to explore different regimes of nonlinear pulse dynamics by tuning the pump frequency. Based on the numerical results that align well with our experimental findings, we propose that varying the pump frequency allows us to transition between shock wave formation and soliton regimes. Hence, we calculate temporal profiles of the photonic field shown in \autoref{fig:3}.  At higher detuning, the system predominantly exhibits shock-wave behavior, characterized by a sharp leading edge and a decaying tail due to self-steepening (\autoref{fig:3}e). As the detuning decreases, we observe a transition where two solitons emerge, driven by the balance between self-phase modulation and group velocity dispersion, leading to a symmetric pulse profile (\autoref{fig:3}c,d).

The waveguide length also plays a critical role. In shorter waveguides, shock waves dominate due to limited interaction length (\autoref{fig:3}h). However, as the waveguide length increases, the system has more space for nonlinear interactions to evolve. This results in the formation of solitons, as oscillations behind the shock front deepen and stabilize into solitonic trains (\autoref{fig:3}i,k). We want to highlight that our system exhibits a strongly nonlinear response, allowing us to observe significant nonlinear effects even in relatively short waveguides with the length of $L = 50 \ \mu$m (as compared to optical fibers), which has the potential to be applied in room-temperature on-chip devices.

\begin{figure}[H]
\centering
\includegraphics[width=1\linewidth]{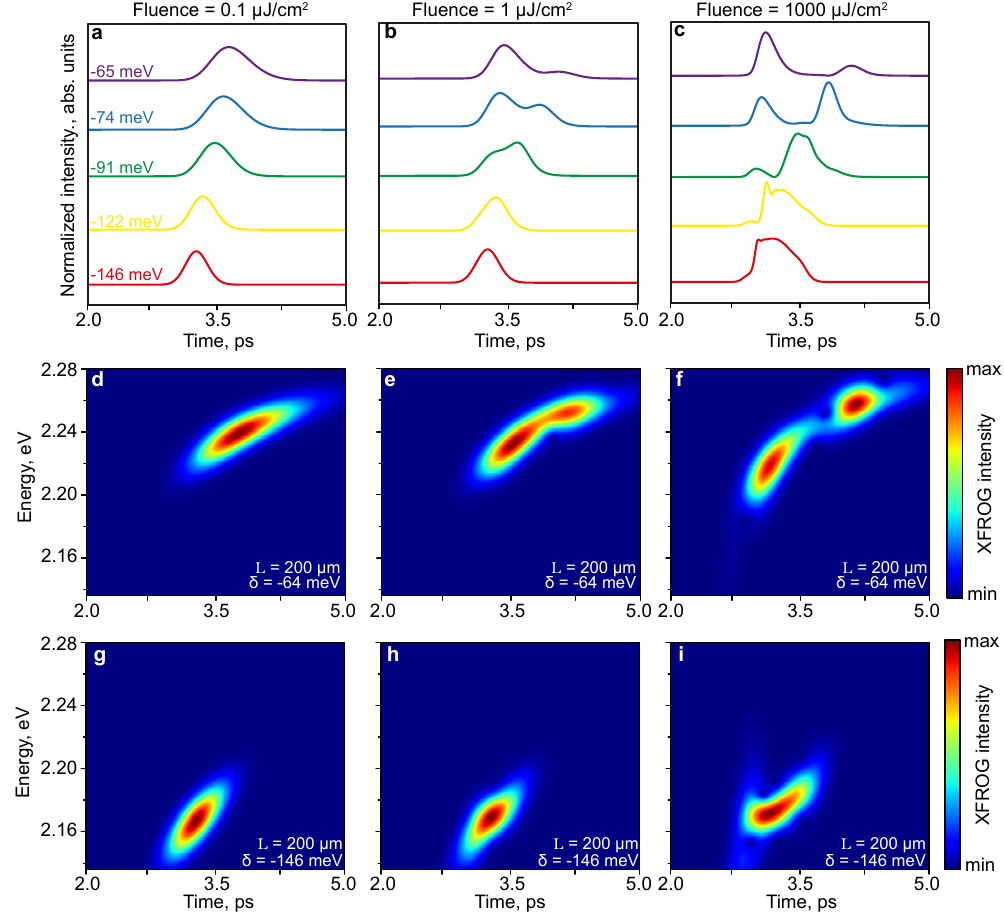}
\caption{{\textbf{Nonlinear effect of the polaritonic waveguide on pulse propagation.}} (\textbf{a-f}) Cross-correlation frequency-resolved optical gating (XFROG) for two detunings and different fluence. (\textbf{c-e})  Simulated transmitted temporal pulse profiles normalized on input pulse plotted for different detunings at three fluences.}
\label{fig:4}
\end{figure}

The discussed dynamics are clearly visible in the modeled temporal profiles shown in \autoref{fig:4}a-c, where decreasing detuning transitions the system from shock wave formation to soliton formation. Specifically, under certain pump conditions (\autoref{fig:4}c), our modeling reveals oscillations trailing the shock front that gradually deepens, eventually forming a pair of solitons. This observation suggests that shock waves and solitons are not isolated phenomena but rather exist on a continuum of behaviors influenced by input parameters such as pump fluence and frequency.

Experimentally, only the spectral domain is accessible. In \autoref{fig:4}d-i, we present the cross-correlation frequency-resolved optical gating (XFROG) data from the model for various input fluences and detunings (see SI for details and different waveguide lengths). XFROG enables the comparison of experimentally obtained spectra with theoretical temporal profiles, capturing the evolution of the transmitted pulse in the spectral-temporal domain. For detuning \(\delta = -64\ \text{meV}\) and a waveguide length of \(L = 200 \ \mu\text{m}\) (\autoref{fig:4}d-f), there is a distinct splitting of the transmitted pulse into two separate peaks with increasing fluence from 0.1 to 1000 $\mu \text{J/cm}^2$, attributed to self-phase modulation. Additionally, the changing slope of this structure indicates variations in group velocity. For detuning \(\delta = -146\ \text{meV}\) and waveguide length \(L = 200 \ \mu\text{m}\) (\autoref{fig:4}g-i), the changing slope is further associated with nonlinear group velocity dispersion, leading to a steep front formation that may represent a shock wave.

\section{Conclusion}

We have revealed a strong effect of SPM and nonlinear GVD in the planar 2D polaritonic waveguide based on halide $\text{MAPbBr}_3$ perovskite at room temperature. The experiments have been carried out by exciting the guided exciton-polaritons using a femtosecond laser resonantly coupled into the waveguide via imprinted grating. 
Due to polariton-polariton interaction, the pulse transmittance exhibits vivid nonlinearity for fluences above 5~$\mu$J$\cdot$cm$^{-2}$. 
The SPM manifests itself in changes in the spectral pulse shape during the pulse propagation through the waveguide, which depends on the incident laser fluence. 
The theoretical analysis has revealed that SPM induces the temporal splitting of the input pulse into two separate pulses and drives the system into the quasi-soliton and shock-waves regimes. Our findings are supported by a detailed experimental study containing wavelength and distance-dependent fs-pulse propagation and the theoretical model that complements experimental data. 

Generally, the strong optical nonlinearity driven by polariton-polariton interactions in perovskite waveguides opens the door to exploring various nonlinear phenomena, such as solitons, shock waves, and self-phase modulation in planar systems. This approach offers a new direction for fabricating integrated on-chip polaritonic devices, making them more reproducible and cost-effective compared to other waveguiding setups that rely on external cavities.

\section{Methods}
\label{ch:Methods}

\textbf{Sample fabrication}

Perovskite waveguide is fabricated utilizing spin-coating and nanoimprint techniques. The first step of thin film fabrication is preparing of perovskite solution.
The perovskite solution is prepared in a glove box nitrogen atmosphere using two salts: methylammonium bromide (MABr) and Lead(II) bromide (PbBr$_2$). The mass of salts is calculated based on the molarity of the solution equal to 0.5, $m_{\text{MABr}} = 55.98 \ \text{mg}$ and $m_{\text{PbBr}} = 183.51 \ \text{mg}$. Salts dissolved in a 1~ml mixture of dimethylformamide (DMF) and dimethylsulfoxide (DMSO), in the ratio DMF:DMSO 3:1.
Before synthesizing the thin film, 12×12~mm$^2$ $\text{SiO}_2$ substrates were cleaned sequentially with soapy water, acetone, and isopropanol. To achieve the higher attribute of surface adhesion, we put substrates to the oxygen plasma cleaning camera for 10 minutes. The spin coating process is carried out in a glove box with a dry nitrogen atmosphere, because of sensitivity to oxygen concentration. The prepared perovskite solution is deposited on top of the substrate and after it accelerates for 3 seconds and rotates at a speed of 3000 rpm for 45 seconds with the spilling of dry chlorobenzene at 25 seconds from the beginning of the spin coating. Thus we obtain a thin film of perovskite $\text{MAPbBr}_3$ with a thickness roughly 110~nm. 

The nanoimprint glass master molds are fabricated by electron beam lithography. At first, the fused silica wafers (1 by 1~cm$^2$) are cleaned by the Hellmanex and DI water solution for 15 minutes in the sonicator, followed by the acetone and isopropanol in the sonicator. Then, those wafers are cleaned with a standard Piranha solution (H$_2$SO$_4$: H$_2$O$_2$ = 3: 1) in the cleanroom. Next, the Piranha cleaned wafers are dried by N$_2$ gas flow and placed on a hot plate before depositing the e-beam resist. Using a spinner, an e-beam resist (PMMA 950 A2) is deposited on wafers to get the target thickness of PMMA for the lift-off procedure, followed by a post-baking of 3 minutes on the hot plate. The fused silica wafer and PMMA are dielectric non-conductors that can accumulate charge during electron beam lithography (EBL) pattering. To avoid the charging problem, a very thin layer (7~nm) of conducting polymer E-Spacer 300Z (SHOWA DENKO) is coated on PMMA using a spinner. Then the wafers are placed in the EBL (FEI Nova NanoSEM 600) chamber. After pattering with a dose value of 250~$\mu$C/cm$^2$, E-Spacer is removed at first by rinsing in DI water for 1 minute. Then the PMMA is developed by the MIBK: IPA solution (1:3) for 1 minute, followed by IPA and DI for 20 seconds respectively. The EBL patterned wafers are placed in the e-beam evaporator (MIDAS e-beam Evaporator) chamber to deposit a 15 nm thick Chromium (Cr). After checking the pattern with the scanning electron beam microscope (SEM) with the Cr, wafers are rinsed in acetone for the lift-off process. A sonicator is also used to accelerate the lift-off process. Then those wafers are dried up with N$_2$ gas flow and placed in the inductively coupled plasma (ICP) (STS Multiplex ICP) chamber to transfer the pattern to fused silica wafer. A mixer of CHF$_3$, SF$_6$, and O$_2$ (20, 20 and, 4 sccm) gas is used to etch down the fused silica where Cr is used as a hard mask. Finally, the unwanted Cr is removed by using chrome etchant and, dry wafers with patterns are stored in a glove box. 

The next stage of fabrication is the optimization of nanoimprint parameters for the fabrication of waveguides based on perovskite thin film.  
It is empirically determined that the necessary pressure exerted for 10 minutes on the synthesised perovskite sample with dimensions of 12 by 12~mm$^2$ is 40~MPa. After that, the synthesised samples are subjected to thermal influence at a temperature of 90~$^\circ$C for 10 minutes. As a result, waveguides with radiation input and output grating elements repeating the geometry of the master mold are obtained. The fabrication scheme of the waveguides is shown in SI. The geometry of the fabricated structures is the inverse geometry of the master mold. Thus, waveguides based on a perovskite thin film with a length in the range of 50 - 200~$\mu$m and a period in the range of 260 - 320~nm with a modulation of 65~nm were fabricated. The couplers and decouplers are characterised using AFM and are shown in Figure S2. It is important to note that master molds can be used multiple times. They must be cleaned with DI water and dried before reuse. 

\textbf{Angle-resolved spectroscopy}

Angle-resolved spectroscopy measurements are performed using a 4f setup featuring a back-focal plane (BFP). This arrangement includes a slit spectrometer connected to an imaging EMCCD camera (Andor Technologies Kymera 328i-B1 + Newton EMCCD DU970P-BVF) and a halogen lamp, which is attached to an optical fibre via a collimation lens for white light illumination. A Mitutoyo Plan Apo Infinity Corrected objective with 10× magnification and a numerical aperture 0.28 is used for excitation and signal collection. Spatial filtering occurs within the intermediate image plane (IP) as it passes through the 4f configuration. Angle-resolved resonant measurements are also conducted using the same setup, employing a femtosecond laser with a tunable wavelength, a repetition rate of 1~kHz, and a pulse width of 200~fs. The laser system includes a mode-locked Ti:sapphire laser operating at 800~nm, which serves as a seed pulse for the regenerative amplifier (Spectra Physics, Spitfire Pro). The wavelength is frequency-doubled to 400~nm using a BBO crystal or combined with an optical parametric amplifier for other pump wavelengths (Light Conversion, OPA). To create a pump spot approximately 50~$\mu$m in size, a lens with a 1000~mm focal length is used to focus the laser beam in the BFP. A CCD camera (Thorlabs 1.6 MP Color CMOS Camera DCC1645C) is used in the collection channel, with a 150~mm tube lens positioned after the beam splitter. This setup allows for imaging in both real and Fourier spaces. To resonantly excite the waveguide mode with a femtosecond laser, we select the appropriate wavelength based on the dispersion of the leaky modes (Figure S4a). The laser spot is positioned on the coupler, and the resonant angle is determined by adjusting the laser’s incident angle in the BFP to match the correct wavevector ($k_x$) corresponding to the dispersion (\autoref{fig:1}c), while simultaneously maximizing the output signal from the decoupler in real space.

\begin{acknowledgement}
HVD gratefully acknowledges support from TUBA — Turkish Academy of Sciences. N.G. thanks Ivan Sinev for the valuable discussion on the visual representation of results. 
\end{acknowledgement}

\begin{suppinfo}

The Supplementary Information includes details on the sample fabrication process with AFM characterization,  scheme of the optical setup, angle-resolved reflectance and transmittance spectra, fitting of polariton modes, coupling efficiency calculations, extended theoretical description, the input pulse shape, and cross-correlation frequency-resolved optical gating analysis for different waveguide lengths.
\end{suppinfo}

\bibliography{Main}

\end{document}


%
\section{Sample fabrication}
\begin{figure}
    \centering
    \includegraphics[width=1\linewidth]{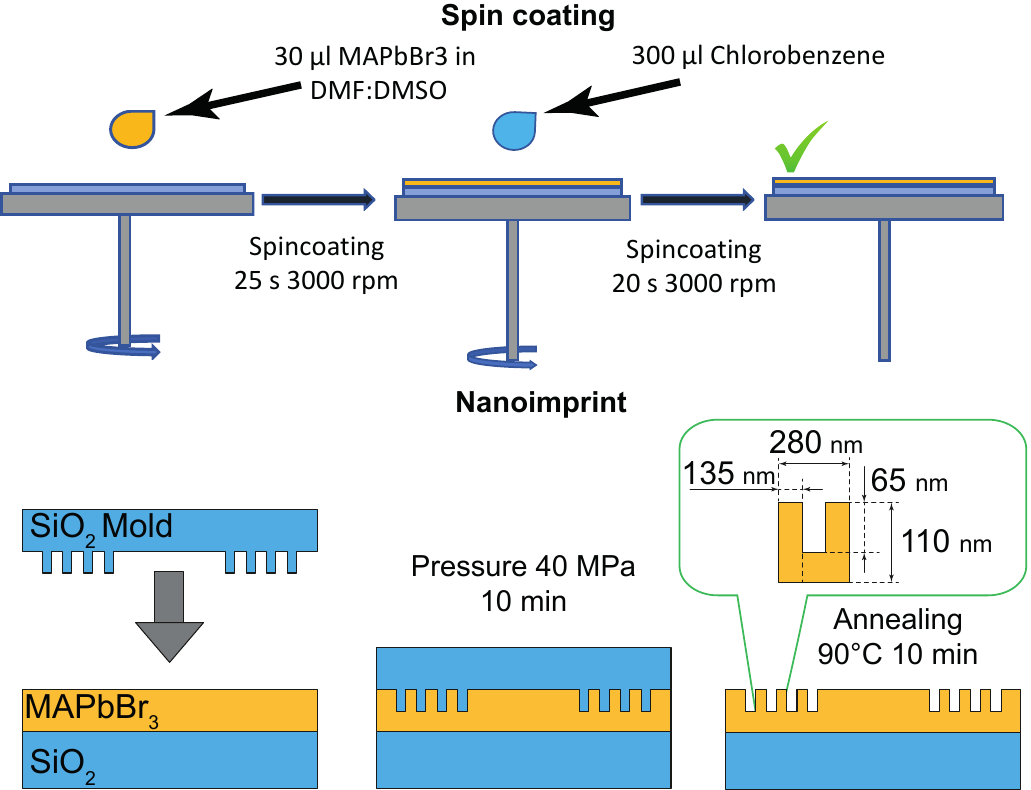}
    \caption{ Scheme of the spin coating and nanoimprint.  }
    \label{fig:S1}
\end{figure}

\begin{figure}
    \centering
    \includegraphics[width=1\linewidth]{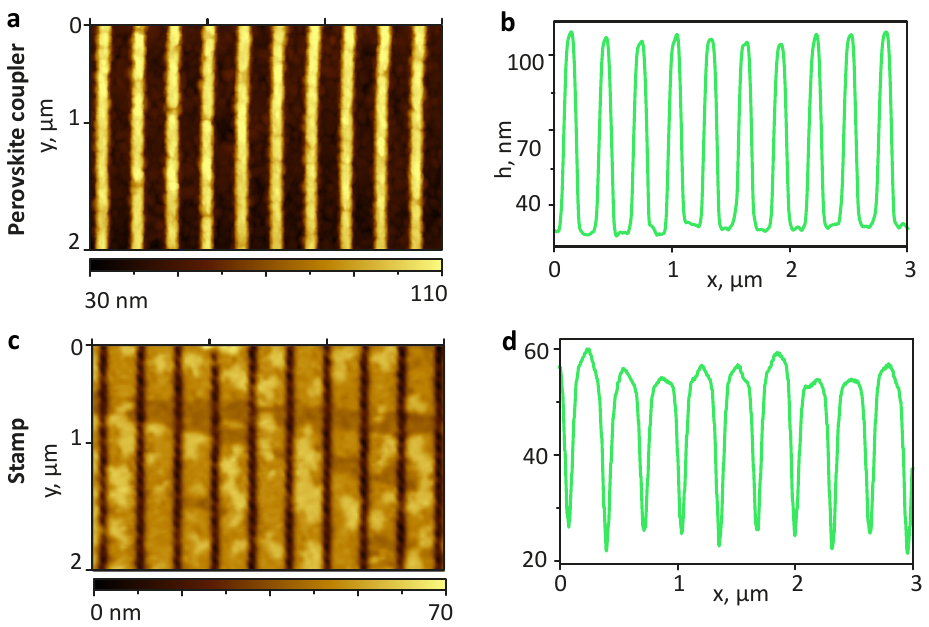}
    \caption{(\textbf{a}) The atomic-force microscopy image of a perovskite coupler and (\textbf{c}) a master mold. (\textbf{b}) The average along the y-axis profile of the perovskite coupler and (\textbf{d}) the coupler.  }
    \label{fig:S2}
\end{figure}

In the main manuscript, we outline the fabrication method for the perovskite waveguide, which involves two primary steps: first, spin coating a thin film of $\text{MAPbBr}_3$, and second, applying nanoimprint lithography using a glass master mold prepared via electron beam lithography. The schematic representation of this fabrication process is provided in \autoref{fig:S1}, illustrating each step in detail from thin film deposition to the pattern transfer using the nanoimprint technique.  
The fabricated gratings used for coupling and outcoupling radiation were characterized using atomic force microscopy (AFM) on the perovskite waveguide system and the corresponding glass master mold. Topographic maps with pseudo-color, which indicate the height variations of the sample, as well as detailed cross-sectional profiles, are shown in \autoref{fig:S2}. These measurements provide a precise visualization of the surface structure and geometrical sizes, offering critical insights into the fabrication quality and surface uniformity of the gratings, which are essential for efficient light coupling and outcoupling. 

\section{Experimental setup}
Here we presented a scheme of the optical setup we use for measuring angle-resolved reflectance and transmittance spectra. The description of the setup is presented in Methods of the main manuscript.  
\begin{figure}
    \centering
    \includegraphics[width=0.8\linewidth]{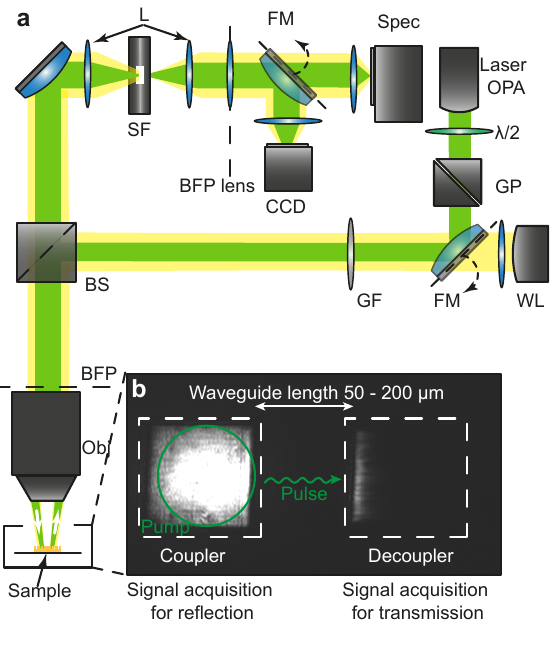}
    \caption{(\textbf{a}) Sketch of the experimental setup used to study the nonlinear response of the perovskite waveguide. (\textbf{b}) The optical image shows light both reflected and transmitted through the waveguide.  }
    \label{fig:S3}
\end{figure}

\section{Measured dispersion of the polaritonic perovskite waveguide}
\begin{figure}
    \centering
    \includegraphics[width=1\linewidth]{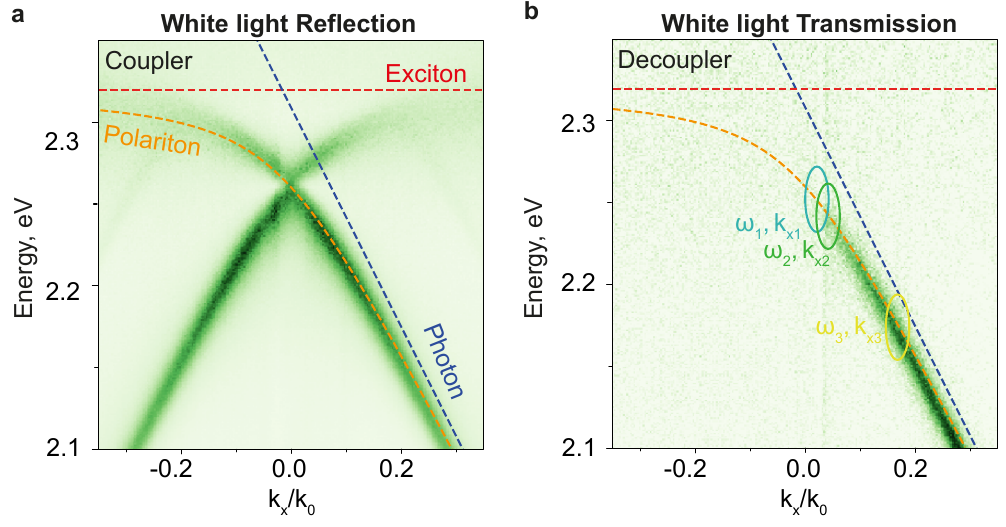}
    \caption{Measured angle-resolved reflectance (\textbf{a}) and transmittance (\textbf{b}) spectrum from coupler and decoupler, respectively, with excitation by white light. Orange lines represent the fitted polariton mode, red lines correspond to the uncoupled exciton resonance, and blue lines correspond to the estimated uncoupled waveguide mode dispersion. Ellipses show polariton states with particular energies and wavevectors excited resonantly using a femtosecond laser. }
    \label{fig:S4}
\end{figure}

We measure angle-resolved reflectance and transmittance spectra (\autoref{fig:S4}) using the optical setup presented above with excitation by white light. The results are described in the main manuscript. 

\section{Fitting by two-coupled oscillator model}

\begin{figure}[H]
    \centering
    \includegraphics[width=0.9\linewidth]{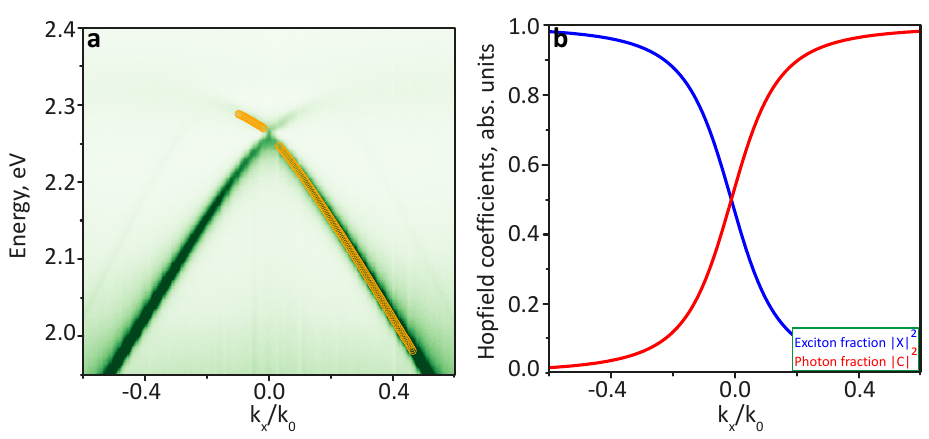}
    \caption{(\textbf{a})  Angle-resolved reflectance spectra with the polariton dispersion extracted by fitting each wavenumber $k_x/k_0$ using a Fano function (indicated by orange circles). (\textbf{b}) Calculated Hopfield coefficients as a function of the normalized wavenumber $k_x/k_0$.}
    \label{fig:S5}
\end{figure}

We extract the polaritonic modes from the experimental data to verify the strong light-matter coupling regime and estimate Hopfield coefficients using the following approach: first, we fit the modes at each $k_x/k_0$ of angle-resolved reflection spectra using a Fano function\cite{miroshnichenko2010fano} to identify the peak positions and spectral width. By compiling the spectral peak positions across all wavenumbers $k_x/k_0$, we derive the experimental mode dispersion (\autoref{fig:S5}a). Since the upper polariton branch (UPB) does not manifest above the exciton resonance due to strong non-radiative absorption, confirmation of strong coupling can only be achieved by fitting the extracted mode to the lower polariton branch (LPB), as predicted by the two-coupled oscillator model \cite{hopfield1958theory}:
\begin{equation}
    E_{LP} = \frac{\widetilde{E}_x + \widetilde{E}_c(k)}{2} - \frac{1}{2}\sqrt{\left(\widetilde{E}_x - \widetilde{E}_c(k)\right)^2 +4 g^2 },
    \label{LPB}
\end{equation}
where $\widetilde{E}_x = E_x - i\gamma_x$ is complex energy accounting for the spectral position and the linewidth of the uncoupled exciton resonance, $\widetilde{E}_c(k) = E_c(k) - i\gamma_c$ is a complex dispersion of the uncoupled cavity photon mode, $g$ - is a light-matter coupling coefficient. The Rabi splitting $\Omega_R$ corresponds to the minimal energy distance between UPB and LPB, however, as UPB does not exist, we can only estimate this value based on the described model:
\begin{equation}
    \Omega_R = \sqrt{4g^2 - (\gamma_c - \gamma_x)^2};
    \label{eq:Rabi}
\end{equation}

The uncoupled cavity photon mode follows a linear dispersion with respect to the wavenumber $k_x/k_0$, since the refractive index is assumed to vary negligibly across the spectral range, excluding the influence of the exciton resonance. Consequently, the uncoupled photon dispersion is approximated as $E_c(k_x) = k\cdot k_x + b$, based on Fourier modal method calculations \cite{li1997new}.
For fitting the LPB, the light-matter coupling coefficient $g$, along with the linewidths of the uncoupled photon $\gamma_c$ and exciton $\gamma_x$, are used as optimization parameters. The resulting real part of the optimized PL dispersion $E_{LP} $ is plotted as the orange dashed curves in \autoref{fig:S4}.
The Hopfield coefficients, which characterize the excitonic and photonic fractions in the polariton, are given by the following expressions:\cite{hopfield1958theory}
\begin{equation} 
 \begin{cases}
    \left| X_{\bm{k}_\parallel} \right|^2 =  \frac{1}{2} \left( 1 + \frac{\Delta E(k_\parallel)}{\sqrt{\Delta E(k_\parallel)^2 + 4g_0^2}} \right),\\
    \left| C_{\bm{k}_\parallel} \right|^2 =  \frac{1}{2} \left( 1 - \frac{\Delta E(k_\parallel)}{\sqrt{\Delta E(k_\parallel)^2 + 4g_0^2}} \right), \\
 \end{cases}
 \label{eq:coeff_hopf}
\end{equation}
We calculate these coefficients using parameters obtained from the fitting procedure described earlier. The dependence of the Hopfield coefficients on the normalized wavenumber $k_x/k_0$ is shown in \autoref{fig:S5}b, providing insight into the evolution of excitonic and photonic contributions across the dispersion.

\section{Coupling efficiency}
To estimate the coupler efficiency, we measure angle-resolved transmission spectra for various waveguide lengths at a fixed detuning of -74 meV. By integrating the transmission signal over both wavelength and wavenumber, we calculate the total transmittance of the system, which includes the coupler, decoupler, and waveguide. 
For a waveguide length of zero, the transmittance of the system reflects the product of the coupling and decoupling efficiencies. By fitting the transmittance dependence on waveguide length with an exponential function, we estimate the coupling efficiency of our system to be 0.168.

It should be noted that, in this analysis, we neglect the nonlinear effects of the coupler and any variation in efficiency due to detuning. We also assume equal coupling and decoupling efficiencies. The waveguide transmittance is then calculated by dividing the total system transmittance (including the couplers) by the square of the coupling efficiency, which accounts for both coupling and decoupling. The input fluence is determined by dividing the incident fluence by the coupling efficiency.   


\section{Linear dependence on input fluence}
\begin{figure}[h!]
    \centering
    \includegraphics[width=0.5\linewidth]{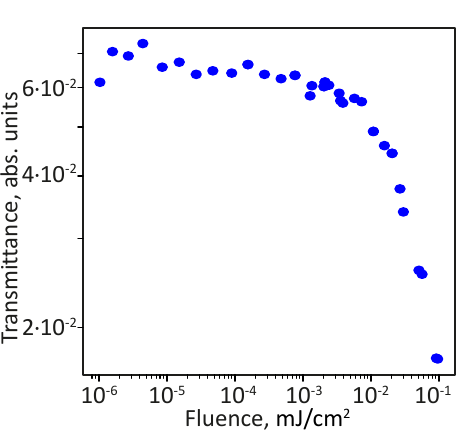}
    \caption{ Transmittance of the waveguide depending on input fluence at the detuning -94 meV with linear regime.}
    \label{fig:S6}
\end{figure}

In the main manuscript, we primarily focus on the nonlinear regime of pulse propagation through the perovskite waveguide. Here, we aim to illustrate the transition from the linear to the nonlinear regime. \autoref{fig:S6} shows the dependence of the waveguide transmittance on input fluence. Initially, the transmittance remains constant, indicating linear behavior. However, a sharp nonlinear decrease in transmittance is observed at approximately 10 $\mu J/cm^2$, marking the onset of the nonlinear regime.     

\section{Extended theoretical model}
Here, we present an in-depth theoretical model. In the main text, we explore the following system of equations to analyze the dynamics involved:

\begin{equation}
    \begin{cases}
        \partial_tA + v_g \partial_xA = -\gamma_{ph}A + i\Omega_R\psi + f_p(x,t),\\
        \partial_t\psi = -(\gamma_{ex} + \gamma_{res})\psi + i\Delta\psi -\Gamma\psi + i\Omega_R A, \label{coh_exciton_eq}\\
        \partial_t\rho  = -\gamma_{\rho}\rho + 2\gamma_{res}|\psi|^2
    \end{cases}
\end{equation}

The first equation is written for the slowly varying amplitude $A$ of the photon-guided mode. We assume that the material and geometrical dispersion of the photonic mode is negligible compared to that appearing because of the interaction between the photons and excitons. Therefore, the dispersion of the photonic component is a straight line with the slope defined by the group velocity $v_g$ at the frequency of the exciton resonance.  The losses in the photonic system are accounted for by the dissipation rate $\gamma_{ph}$. The photonic component is excited by the driving force (currents) $f_p$ produced by the incident pulse in the area of the input coupler.

The photonic subsystem is coupled to the excitonic one with the coupling strength characterized by Rabi splitting $\Omega_R$.  The coherent excitons are described by the slowly varying envelope amplitude $\psi$. We assume that the excitons do not interact with their neighbors linearly, and therefore, Eq.~(\ref{coh_exciton_eq}) does not contain a spatial derivative of $\psi$. In terms of dispersion characteristic, this means that the dispersion of the exciton is a horizontal straight line at the frequency of the exciton resonance. The losses in the exciton sub-system are defined by the total dissipation rate $\gamma_{ex}+\gamma_{res}$ consisting of the excitons decay accounted by $\gamma_{ex}$ and by the rate of the scattering to the reservoir of the incoherent excitons accounted by $\gamma_{res}$.

Let us remark that the non-interacting photons and excitons are dispersionless, however, in the so-called strong coupling regime, the photon-exciton interaction opens the gap in the dispersion characteristic and the hybrid excitations containing both photon and exciton component experiences strong dispersion in the vicinity of the exciton resonance. The dispersion of these hybrid excitations (polaritons) is given by
\begin{equation}
\omega=\frac{ v_g k+i(\gamma_{ph}+\gamma_{ex}+\gamma_{res}) \pm\sqrt{\left(v_g k+i(\gamma_{ph}-\gamma_{ex}-\gamma_{res}) \right)^2+4\Omega_R^2  }   }{2}. \label{dispersion}
\end{equation}
A strong coupling regime is defined as one where at zero detuning from the resonance, the real part of the eigenfrequencies of the modes are different (there is a gap between the dispersion characteristic, and thus, there are the upper and the lower polariton branches).

To reproduce the dynamics observed in the experiments, it is necessary to account for the reservoir of long-living incoherent excitons. This reservoir is characterized by the incoherent exciton density $\rho$. The scattering from the coherent excitons to the incoherent ones is $2 \gamma_{res} |\psi|^2$; it is assumed that the incoherent excitons can never become the coherent ones, and so for the coherent exciton, the scattering to the incoherent ones is just an additional contribution to the effective absorption. The lifetime of the coherent excitons is defined by the coefficient $\gamma_{\rho}$. For calculations we use $\gamma_{res} = 5.71$ meV and  $\gamma_{\rho} = 0.33$ meV.

Typically, the dependency of the frequency and the losses of the photon mode on its intensity is negligible, and therefore, the equation for $A$ is linear. At the same time, the densities of the excitons needed to observe the blueshift of the exciton resonance are easily achievable in the experiments. The non-radiative decay of the coherent excitons can also depend on the exciton density. It is important that both the coherent and incoherent excitons can contribute to the nonlinear exciton frequency shift and to nonlinear losses in the exciton subsystem. These nonlinear dependencies of the resonant frequency and the losses are accounted for by $\Delta$ and $\Gamma$, which are functions of the exciton densities. In our numerical modeling, we assume that the exciton frequency blueshift and the losses saturate at some level, and thus, we phenomenologically approximate these dependencies as 
 $$\Delta =\delta_0 \frac{\tilde \rho}{I_{\Delta} +\tilde \rho},$$
$$\Gamma =\Gamma_0 \frac{\tilde \rho}{I_{\Gamma} +\tilde \rho},$$
where $\tilde \rho = |\psi|^2+\rho$ is total density of the coherent and incoherent excitons. The parameters $I_{\delta}$ and $I_{\Gamma}$ define the saturation levels  and $\delta_0$ and $\Gamma_0$ - the strengths of the nonlinearities.

The parameters of the nonlinearities can be estimated by fitting the known values of the cubic and fifth-order nonlinearities. For this, we expand the saturable nonlinearity in the Taylor series and then equalize the coefficients in front of the first and the second terms. The expansions read $\Delta=\frac{\delta_0}{I_{\Delta}} \tilde \rho - \frac{\delta_0}{I_{\Delta}^2} \tilde \rho^2$ and $\Gamma=\frac{\Gamma_0}{I_{\Gamma}} \tilde \rho - \frac{\Gamma_0}{I_{\Gamma}^2} \tilde \rho^2$. Knowing the cubic $\alpha_3$, $\beta_3$ and quintic $\alpha_5$, $\beta_5$ nonlinearities coefficients we obtain $I_{\Delta}=-\frac{\alpha_3}{\alpha_5}$, $\delta_0=-\frac{\alpha_3^2}{\alpha_5}$ and $I_{\Gamma}=-\frac{\beta_3}{\beta_5}$, $\Gamma_0=-\frac{\beta_3^2}{\beta_5}$. In our theoretical calculations, we use parameters $\alpha_3 = V_{XX} = 0.02 \ \mu\text{eV}\cdot\mu\text{m}^3$  and $\alpha_5 = V_{XX2} =-3\cdot 10^{-9} \ \mu\text{eV}\cdot\mu\text{m}^6$, $\beta_3= 5 \cdot 10^{-3} \ \mu\text{eV}\cdot\mu\text{m}^3$ and $\beta_5 = -3.1\cdot 10^{-16}  \ \mu\text{eV}\cdot\mu\text{m}^6 \approx 0$, corresponding closely to values from previous work \cite{masharin2023room}, with minor variations.

\section{Input pulse shape}

We cannot directly measure the exact shape of the pulse as it couples into the waveguide via the coupler. Therefore, in our simulations, we assume a Gaussian temporal profile with an added chirp to account for the substantial effects of group velocity dispersion: 
\begin{equation}
    f_p(x,t) = \sqrt{\frac{1}{1+2i\frac{\Theta_2 x}{T_0^2}}}\text{exp}\left(-\left(\frac{x-x_0}{w_0}\right)^2\right)\text{exp}\left(-\frac{(t-t_0)^2\left(1-2i\frac{\Theta_2 x}{T_0^2}\right)}{T_0^2\left(1 + 4i\frac{\Theta_2^2 x^2}{T_0^4}\right)}\right)\text{exp}\left(i\omega_p t - ik_p x\right), 
\end{equation}
where $\Theta_2 = \frac{\partial^2 k}{\partial \omega^2}$ - chirp, $w_0 $ - spatial aperture, $T_0 $ - pulse duration.
The parameters of the Gaussian profile are carefully optimized by comparing simulated results to experimental data, focusing on matching key characteristics such as spectral bandwidth, peak positions, and splitting dynamics. The use of a positive $\Theta_2$ in our simulations corresponds to normal dispersion, where higher-frequency components travel faster than lower-frequency ones.

This approach offers an effective approximation of the coupled pulse, capturing the dominant physical effects that govern its evolution within the waveguide. However, discrepancies between the calculated and experimental data may still arise due to mismatches in the assumed input pulse shape.

\section{Fitting transmitted spectra by the sum of two Gauss functions}
\begin{figure}[h!]
    \centering
    \includegraphics{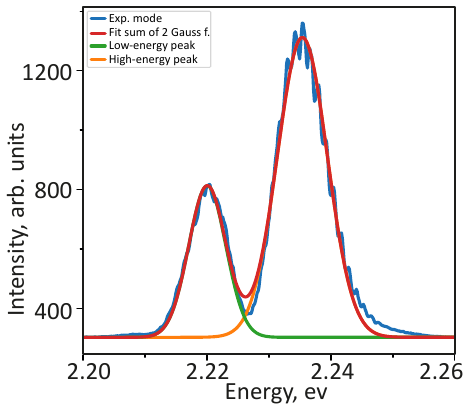}
    \caption{ Integrated intensity of the transmitted spectra through a waveguide of length $L=200$~$\mu$m at a detuning of $\delta = -74$~meV (blue line). The red line represents the fitted sum of two Gaussian functions to the experimental data. The individual Gaussian components, corresponding to low-frequency (green line) and high-frequency (orange line) peaks, are shown separately.}
    \label{fig:S7}
\end{figure}
We observe two distinct peaks in the temporal domain and spectral domain. Hence, we fit integrated over angles spectra by the sum of two Gauss functions to estimate the parameters of the transmitted pulses:
\begin{equation}
    f(\omega) = \frac{A_1}{\sqrt{2\pi}\sigma_1} \text{exp}\left(-{\frac{(\omega - \omega_{0_1})^2}{2\sigma_1^2}}\right) + \frac{A_2}{\sqrt{2\pi}\sigma_2} \text{exp}\left(-{\frac{(\omega - \omega_{0_2})^2}{2\sigma_2^2}}\right) + C,
\label{eq:gausses}
\end{equation}
where $\omega_{0_1}, \omega_{0_2}$ represent the central frequencies of the peaks, $\sigma_1, \sigma_2$ -- denote the peak dispersions (with the full width at half maximum given by $\Delta\omega = 2\sqrt{2\text{ln}2}\sigma$), $A_1, A_2$ are the peak amplitudes, and $C$ is a background constant. All these parameters are determined by fitting the $k_x/k_0$ integrated spectra using the sum of two Gaussian functions spectra (\autoref{eq:gausses}). \autoref{fig:S7} illustrates the fitting results, showing two distinct peaks within the pulse structure. The experimental mode is accurately fitted by the sum of two Gaussian functions (\autoref{eq:gausses}). The green and orange lines represent the low-energy and high-energy peaks, respectively. These peaks closely match the experimental data in terms of amplitude, peak position, and frequency, enabling precise estimation of the spectral feature dependence on input fluence shown in the main text.

    \section{Cross-correlation frequency-resolved optical gating (XFROG)}
To gain deeper insight into the temporal and spectral characteristics of the pulse, we performed cross-correlation frequency-resolved optical gating (XFROG) calculations for different detunings and waveguide lengths based on temporal profiles of the filed obtained from the theoretical calculations $A(t)$. We compute the Fourier transform of the product of the photonic field's temporal profile, $ A(t)$, and a Gaussian-shaped probe pulse, $R(t, \tau) = \text{exp}\left(-\frac{(t-\tau)^2}{w_{ref}^2}\right)$, for each position of the probe pulse \(\tau\). The resulting quantity is given by:

\begin{equation}
    F(\omega, \tau) = \left|\int_{-\infty}^\infty A(t) R(t, \tau) e^{-i\omega t} \, dt\right|^2
\end{equation}
Finally, we plot $ F(\omega, \tau) $ as a function of frequency $\omega$ and time $\tau$. XFROG is a powerful technique that allows us to resolve the pulse shape in the time domain while simultaneously providing spectral information. This dual resolution helps us to establish a clear correspondence between the peaks observed in the spectral and temporal domains.

\begin{figure}[h!]
    \centering
    \includegraphics[width=1\linewidth]{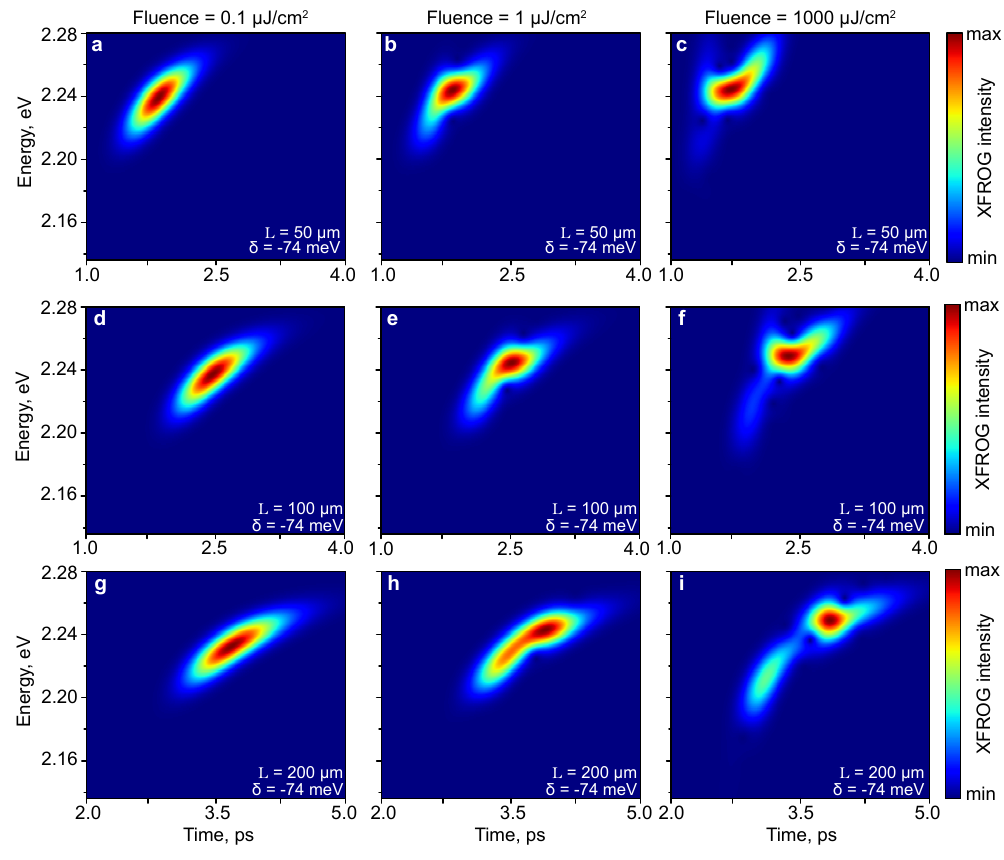}
    \caption{ (\textbf{a-i}) Cross-correlation frequency-resolved optical gating (XFROG) for different waveguide lengths and different fluence.}
    \label{fig:S8}
\end{figure}

By applying XFROG to our system, we can differentiate distinct peaks in the time domain that were previously difficult to distinguish using spectral data alone. For each detuning and waveguide length, the XFROG analysis allowed us to identify which peaks in the spectral domain corresponded to those in the temporal domain. The results from the XFROG analysis are particularly useful for interpreting the experimental data, as they reveal the underlying temporal structure of the pulse that is not directly observable in the spectral measurements. By resolving these relationships, we can more accurately track the transition from shock waves to solitons, as well as understand how changes in detuning and waveguide lengths impact the overall pulse dynamics.

\bibliography{achemso-demo}